\documentstyle[prb,aps,psfig]{revtex}

\begin{document}
\draft

\twocolumn[\hsize\textwidth\columnwidth\hsize\csname
@twocolumnfalse\endcsname

\title{Dynamical properties of liquid lithium 
above melting.} 
\author{J. Casas, D. J. Gonz\'alez and L. E. Gonz\'alez} 
\address{Departamento de F\'\i sica Te\'orica, Universidad de Valladolid,
47011 Valladolid, SPAIN} 
\date{\today}

\maketitle

\begin{abstract}
The dynamical properties of liquid lithium at several thermodynamic 
states near the triple point have been studied within the framework 
of the mode-coupling theory. We present a 
self-consistent scheme which, starting from the knowledge of the    
static structural properties of the liquid system, allows the theoretical 
calculation of several single particle 
and collective dynamical properties. 
The study is complemented by performing Molecular Dynamics simulations 
and the obtained results are compared with the available 
experimental data. 
\end{abstract}

\pacs{ {\bf PACS:} 61.20.Gy; 61.20.Lc; 61.25.Mv }

]

\narrowtext

\section{Introduction.}

In the last two decades, considerable progress has been achieved, both 
experimental and theoretical, in the understanding of the dynamical 
properties of liquid metals, and in particular the alkali metals because 
of their status as paradigms of the so called ``simple liquids''. 
On the experimental side, and restricting to the liquid alkali metals, nearly 
all elements of this group have been studied by neutron scattering 
experiments: Li \cite{Jong1,Jong2,Jong3}, Na \cite{MorGla},
Rb \cite{Pilg} and Cs \cite{MorBo}. Moreover, liquid Li has also been 
investigated by means of  inelastic X-ray scattering 
experiments \cite{Seldmeier,Burkel,Sinn}.
Closely linked to this progress, has been the development of the 
computer simulation techniques because of 
its ability to go beyond the scattering experiments, as they allow to 
determine certain time correlation functions which are not accesible in 
a real experiment. Several computer simulation studies have already been 
performed, mainly for alkali  
metals \cite{npamd2,Shimojo,Kahl,BaluTorVall1,BaluTorVall2}, 
although other systems have also 
been studied, e.g., the liquid alkaline-earths \cite{Alemany} or liquid
lead \cite{Gudowski}.

On the theoretical side, this progress can be linked with the development
of microscopic theories which provide a better understanding of the physical
mechanisms ocurring behind the dynamics of simple liquids \cite{Gotzext}. It
was quite important to notice that the decay of several time-dependent
properties can be explained by the interplay of two different dynamical
processes \cite{Sjobook,Sjothself,SjoJPC,Sjothcoll,SjoRb,Balubook}.  
The first one, which leads to a rapid initial decay, is due to
the effects of fast, uncorrelated, short range interactions (collisional
effects) which can be broadly identified with ``binary'' collisions. The
second process, which usually leads to a long-time tail, can be attributed
to the non-linear couplings of the dynamical property of interest with slowly
varying collective variables (``modes'') as for instance, density
fluctuations, currents, etc., and it is referred to as a mode-coupling
process. Although the resulting analytical 
expressions are rather 
complicated, it is sometimes possible to make some simplifying 
approximations leading to more tractable expressions while retaining 
the essential physical features of the process. In this way,  
when dealing with liquids close to the triple point or supercooled liquids,
the relevant coupling modes are the density fluctuations, and therefore one 
can ignore the coupling to currents, leading to much simpler expressions.  
Sj\"ogren \cite{SjoJPC,SjoRb} first applied this theory, with 
several simplifications, to 
calculate the velocity autocorrelation function and its memory function, in 
liquid argon and rubidium at thermodynamic conditions near their triple 
points, and obtained results in qualitative agreement with 
the corresponding Molecular Dynamics (MD) computer simulations.
Also, it is worth mentioning the theoretical studies  
carried out by Balucani and coworkers 
\cite{BaluTorVall1,BaluTorVall2,BaluD} for the liquid alkali metals close 
to their triple points, which basically focused on  
some transport properties, such as the self-diffusion coefficient and the 
shear viscosity, as well as other single particle 
dynamical properties (i.e., velocity autocorrelation function, its memory 
function and mean square displacement). From their work, which 
included several 
simplifications into the mode-coupling theory, they asserted that the 
single particle dynamics of the alkali metals is, at least 
qualitatively, well described by this theory.
Based on those ideas, we 
have recently developed a theoretical approach \cite{GGCan} that
allows a self-consistent calculation of all the above transport and 
single particle dynamical properties, and which has been applied to liquid
lithium \cite{GGCan} and the liquid alkaline-earths \cite{Alemany} near
their triple points, leading to theoretical results in reasonable agreement
with simulations and experiment. 

At this point, it must be stressed that within the mode-coupling theory, 
both the single particle and the collective dynamical magnitudes 
are closely interwoven and therefore, the 
application of this theory to any liquid system should imply 
the self-consistent solution of the analytical expressions appearing in 
the theory. However, none of the above mentioned theoretical 
calculations have 
been performed in this way; in fact, the usual practice 
is to take the input dynamical magnitudes, needed for the evaluation of
the mode-coupling expressions, either from MD simulations
or from some other theoretical approximations. 
As a clear example of this procedure let us consider the intermediate 
scattering function, $F(k,t)$. 
This is a central magnitude within the mode-coupling theory, which appears 
in the determination of several single particle and collective dynamical 
properties.     
However, we are not aware of any self-consistent 
theoretical calculation for this magnitude, performed either  
within the mode-coupling theory or some simplified version of it. 
Up to our knowledge, $F(k,t)$ has always been obtained either
from MD simulations \cite{Shimojo,Gudowski}, or
within some approximation, in particular, the viscoelastic 
model or some simplification of it 
\cite{BaluTorVall1,BaluTorVall2,Alemany,BaluD,GGCan,TorBalVer}.

Furthermore, the application of the self-consistent scheme of Ref.
\onlinecite{GGCan} to lithium and the alkaline-earths, showed some
deficiencies in the calculated magnitudes, which were attributed to the 
inaccuracies of the viscoelastic intermediate scattering functions,
and we suggested that theoretical efforts should be directed to obtain a
more accurate description of $F(k,t)$.

In this work we present a first step in this direction, by
proposing a self-consistent determination, within the mode-coupling theory, 
of the intermediate scattering function. 
This is performed through a theoretical framework which generalizes our 
previous self-consistent approach  
\cite{GGCan} in order to include both collective and single particle  
dynamical properties. 
This new procedure is then applied to study the dynamical 
properties of liquid 
lithium at several thermodynamic states above the triple point, 
for which there are recent experimental 
inelastic neutron \cite{Jong1,Jong2,Jong3} and 
X-ray \cite{Sinn} scattering data available.

The study of liquid Li from
a fundamental level poses a number of noticeable problems. For instance, the
experimental determination of its static structure shows specific problems
related to the small mass of the ion, and to the possible importance of
electron-delocalization effects \cite{Lugt}. Moreover, the determination of
an accurate effective interionic pair potential is not easy, and several
pseudopotentials have been proposed \cite{npavmhnc}. Theoretical
calculations, using the Variational Modified Hypernetted Chain (VMHNC)
theory of liquids \cite{npavmhnc}, as well as MD 
simulations \cite{npamd} have shown that a good description of the 
equilibrium properties of liquid lithium can be obtained by using an 
interionic pair potential with no adjustable parameters derived from 
the Neutral Pseudo Atom (NPA) method \cite{npavmhnc}.

In all theories for the dynamic structure of liquids, the equilibrium
static properties are assumed to be known a priori. By using the
NPA to obtain the interionic pair potential, and the VMHNC to obtain the
liquid static structure, we are able to obtain these input data,
needed for the calculation of the dynamical properties, from the
only knowledge of the atomic number of the system and its thermodynamic
state.

The paper is organized as follows. In section \ref{theory} we describe the
theory used for the calculation of 
the dynamical properties of the system and we 
propose a self-consistent 
scheme for the evaluation of several single particle and collective 
properties.  
Section \ref{simul} gives a brief 
description of the computer simulations and in section \ref{litio} we 
present the results obtained when this theory is applied to liquid 
lithium at several thermodynamic states above the triple point. 
Finally we sum up 
and discuss our results.

\section{Theory.}

\label{theory}

\subsection{Collective Dynamics.}

The dynamical properties of liquid lithium have been 
measured by means of inelastic neutron scattering (INS) 
\cite{Jong1,Jong2,Jong3} 
at T=470, 526 and 574 K and also by inelastic X-ray scattering (IXS) 
\cite{Seldmeier,Sinn} 
at T=488 K and T=533 K. Whereas the 
IXS experiment provides information on the collective dynamics of the system, 
i.e., those properties described by the dynamic structure factor 
$S(k, \omega)$, 
the INS experiment allows also to obtain information about the single 
particle dynamics, which are enclosed in the total 
dynamic structure factor $S_{\rm tot}(k, \omega)$ 

\begin{equation}
\label{Stot}
S_{\rm tot}(k,\omega) = \frac{\sigma_c}{\sigma_c+\sigma_i} S(k,\omega) + 
\frac{%
\sigma_i}{\sigma_c+\sigma_i} S_s(k,\omega) \, , 
\end{equation}

\noindent where $S_s(k, \omega)$ is the self dynamic structure factor and 
$\sigma_i$ and $\sigma_c$ denote the incoherent and coherent neutron 
scattering cross sections respectively, which in the present case take on 
the following values: 
$\sigma_i(^7$Li$) = 0.68$ barns and 
$\sigma_c(^7$Li$) = 0.619$ barns.  
The dynamic structure factor may be obtained as 

\begin{equation}
S(k, \omega) = \frac{1}{\pi} \; Re \; \tilde{F}(k, z= -i\omega) \, ,
\end{equation}

\noindent where $Re$ stands for the real part and 
$\tilde{F}(k, z)$ is the Laplace transform of the  
intermediate scattering function, $F(k, t)$, i.e., 

\begin{equation}
\label{LapF}
\tilde{F}(k, z)= \int_0^{\infty} \; dt \; e^{-zt} F(k, t) \, .
\end{equation}

\noindent Resorting to the memory function formalism, $\tilde{F}(k,z)$ 
can be expressed as \cite{SjoJPC}
 
\begin{equation}
\label{MfF}
\tilde{F}(k, z) = S(k)\left[ z + \frac{\Omega^2(k)}{z 
+\tilde{\Gamma}(k, z)}\right]^{-1} \, ,
\end{equation}

\noindent where $\tilde{\Gamma}(k, z)$ is the Laplace transform of the 
second-order memory function $\Gamma(k, t)$, and 

\begin{equation}
\Omega^2(k)=k^2/\beta m S(k)  \, ,
\end{equation}

\noindent where $m$ is the mass of the particles, $\beta$ is the inverse 
temperature times the Boltzmann constant and $S(k)$ is the 
static structure factor of the liquid. We note that as we 
consider spherically symmetric potentials and homogenous systems, 
the dynamical magnitudes to be studied here will only depend on the 
modulus $ k = \mid \vec{k} \mid $.

In simple liquids, the reason for dealing with memory functions, is 
connected with the development of microscopic 
theories where these dynamical magnitudes play a key role 
\cite{Sjobook,Balubook}. The time decay of 
these magnitudes is explained by the interplay of two 
different dynamical processes: 
one, with a rapid initial decay, due to
fast, uncorrelated short range interactions, and a 
second process (known as a mode-coupling process), 
with a long-time tail, due  
to the non-linear couplings of the specific dynamical magnitude 
with slowly varying collective variables (``modes'') as, for instance, 
density fluctuations, currents, etc.  
Therefore, according to these ideas,  
$\Gamma(k, t)$ is now descomposed  
as follows \cite{Sjothcoll,SjoRb,Balubook}  
 
\begin{equation}
\label{Gammatot}
\Gamma(k, t) = \Gamma_B(k, t) + \Gamma_{\rm MC}(k, t) \, ,
\end{equation}
 
\noindent where 
the first term, $\Gamma_B(k, t)$, has a fast time dependence,
and its initial value and curvature can be related 
to the static  structure of the system.
The second term, the mode-coupling 
contribution $\Gamma_{\rm MC}(k, t)$, aims to take into account 
repeated correlated 
collisions, starts as $t^4$, grows to a maximum and then decays rather 
slowly. We now briefly describe both terms and for more details we 
refer the reader to Ref. 
\onlinecite{Sjobook,Sjothself,SjoJPC,Sjothcoll,SjoRb,Balubook}.

\subsubsection{The binary-collision term.} 
At very short times, the 
memory function is well described by 
$\Gamma_B(k, t)$ only; moreover, both 
$\Gamma(k, t)$ and $\Gamma_B(k, t)$ have the same initial value and
curvature, which are determined by the first six moments of the dynamic
structure factor \cite{Balubook}. In particular, we have

\begin{equation}
\label{Gambin}
\Gamma(k,0)= \Gamma_B(k,0)= \frac{3 k^2}{\beta m} + \Omega_0^2 + 
\gamma_d^l(k) - \Omega^2(k) \, ,
\end{equation}

\noindent where  
$\Omega_0$ is the Einstein frequency and 
$\gamma_d^l(k)$ stands for the second moment of 
the longitudinal current correlation function; their respective 
expressions are 

\begin{equation}
\Omega_0^2 = \frac{\rho}{3 m} \int d\vec{r} \, \, g(r)\, 
{\nabla}^2 \varphi (r) \, , 
\end{equation}

\begin{equation}
\gamma_d^l(k) = - \frac{\rho}{m} \int d\vec{r} \, 
e^{-i \vec{k} \cdot \vec{r}} 
g(r) \, (\hat{k} \cdot \vec{\nabla})^2 \varphi(r) \, ,
\end{equation}

\noindent with $\varphi (r)$ and $g(r)$ denoting respectively the 
interatomic pair potential and the pair distribution function of the liquid 
system with number density $\rho$, and $\hat{k}=\vec{k}/k$. 

The binary term includes all the contributions to $\Gamma(k, t)$ to 
order $t^2$. Moreover, as 
the detailed features of the ``binary'' dynamics of systems with continuous
interatomic potentials are rather poorly known, we resort 
to a semi-phenomenological
approximation that reproduces the correct 
short time expansion. Although several functions could 
fulfill this requirement, 
the most widely used in the literature are the squared hyperbolic secant and 
the gaussian. Although both start in the same way, their tails are different 
with the gaussian being somewhat narrower. Here, for reasons to be 
discussed below, we use the gaussian expression, i.e., 

\begin{equation}
\label{gammaB}
\Gamma_B(k, t)= \Gamma_B(k,0)  
\; e^{-t^2/\tau_l^2(k)} \, ,
\end{equation}

\noindent where $\tau_l(k)$ is a relaxation time, whose form can be    
determined from a short time expansion of the formally 
exact expression of the binary term, and the result is related to 
the sixth moment of $S(k, \omega)$. In this way, 
after making the superposition approximation for the three-particle 
distribution function, it is obtained 
\cite{Balubook}

\begin{eqnarray}
\label{Tauele}
& & \frac{\Gamma_B(k,0)}{\tau_l^2(k)}  = \nonumber \\
& & \frac{3 k^2}{2 \beta m} 
\left[ \frac{2 k^2}{\beta m} 
+ 3 \Omega_0^2 +  
2 \gamma_d^l(k)\right] + \nonumber \\
& & (\frac{3 \rho}{\beta m^2})\, i k \int d \vec{r}\, 
e^{-i \vec{k} \cdot  \vec{r}} 
g(r)(\hat{k} \cdot \vec{\nabla})^3 \varphi(r) +
\nonumber \\  
& & \frac{\rho}{m^2} \int d \vec{r}\, [1 - e^{-i\vec{k}\cdot\vec{r}}] \,
[{\hat{k}}^{\alpha}\nabla^{\alpha}\nabla^{\gamma}\varphi(r)]\, g(r) \,
[{\hat{k}}^{\beta}\nabla^{\beta}\nabla^{\gamma}\varphi(r)] + \nonumber \\ 
& & \frac{1}{2 \rho} \int \frac{d {\vec{k}}^{\prime}}{(2 \pi)^3} 
{\hat k}^{\alpha} \gamma_d^{\alpha \gamma}({k}^{ \prime}) 
\left\{ \left[ S({k}^{ \prime}) -1 \right] + \left[S(\mid \vec{k}  - 
\vec{{k}^{ \prime}}\mid )-1 \right] \right\} 
\nonumber \\
& & \times \left\{ \gamma_d^{\beta \gamma}({k}^{ \prime})-
\gamma_d^{\beta \gamma}
(\mid \vec{k}-\vec{{k}^{ \prime}}\mid ) \right\} {\hat{k}}^{\beta} \, ,
\nonumber \\
& & 
\end{eqnarray}

\noindent where summation over repeated indices is 
implied ($\alpha, \beta, \gamma = x, y, z$). Here 

\begin{equation}
\gamma_d^{\alpha \beta}(k) = - \frac{\rho}{m} \int d \vec{r} \,\, 
e^{-i \vec{k} \cdot \vec{r}}  \, g(r) \, \nabla^{\alpha} \nabla^{\beta} 
\varphi (r) \, ,
\end{equation}

\noindent and therefore, the relaxation time can be evaluated from the 
knowledge of the interatomic pair potential and its 
derivatives as well as the 
static structural functions of the liquid system.




\subsubsection{The mode-coupling component.} 
The inclusion of a slowly decaying time 
tail is a basic ingredient in order to obtain, at least, a qualitative  
description of $\Gamma(k, t)$. A rigorous treatment of this term implies the 
combined use of kinetic and mode-coupling 
theories.  
In principle, coupling to several modes should be
considered (density-density coupling, density-longitudinal 
current coupling and density-transversal current coupling)
but for thermodynamic conditions 
near the triple point the most important contribution
arises from the density-density coupling. 
So, restricting the mode-coupling 
component to the density-density coupling term only, 
this contribution can be written as \cite{SjoRb}

\begin{eqnarray}
\label{gammamc}
& &  \Gamma_{\rm MC}(k, t)  = 
\frac{\rho}{\beta m} \int \frac{d\vec{k^{\prime}}}
{(2 \pi)^3}\, \hat{k} \cdot \vec{k^{\prime}} \, c(k^{\prime}) \nonumber \\
& & \times
\left[ \hat{k} \cdot \vec{k^{\prime}} \, c(k^{\prime}) + 
\hat{k} \cdot (\vec{k}-\vec{k^{\prime}}) \, 
c(\mid k-k^{\prime} \mid)\right] \nonumber \\ 
& & \times \left[F( \mid \vec{k} - \vec{k^{\prime}} \mid, t) \, 
F(k^{\prime}, t) - 
F_B(\mid \vec{k} - \vec{k^{\prime}} \mid, t) \, F_B(k^{\prime}, t)\right]
\, ,
\nonumber \\
& & 
\end{eqnarray}

\noindent where   
$F_B(k^{\prime}, t)$ denotes the binary part of 
the intermediate scattering function, $F (k^{\prime}, t )$. We note that the 
part involving the product of the $F_B(k, t)$'s has the effect of making 
$\Gamma(k, t)$ very small at short times; therefore the influence of the 
approximation made for the $F_B(k, t)$ is negligible beyond a short time 
interval, because the intermediate and long time features of 
the $\Gamma(k, t)$ are controlled by the term involving 
the product of the $F(k, t)$'s.  
Following Sj\"ogren \cite{SjoRb}, we assume that the ratio 
between $F(k, t)$ and 
its binary part can be approximated by the ratio between their corresponding 
self-parts, i.e.,  

\begin{equation}
\label{FsB}
F_B(k, t) =\frac{F_{sB}(k, t)}{F_s(k, t)} \,  
F(k, t) \, ,
\end{equation}

\noindent where $F_s(k, t)$ is the self intermediate scattering function 
and $F_{sB}(k, t)$ stands for its binary part, which is now approximated 
by the ideal gas expression   

\begin{equation}
F_{sB}(k, t) = F_0(k, t) \equiv \exp [-\frac{1}{2m \beta}k^2t^2] \, . 
\end{equation}

Now, once $F_s(k, t)$ is specified, the self-consistent evaluation 
of the above formalism will yield to $F(k,t)$, and, through 
them, to all the collective dynamical properties discussed 
so far.
An appropiate expression for $F_s(k, t)$ is provided by the gaussian 
approximation

\begin{equation}
\label{FsuS}
F_s(k,t) = \exp [-(\frac{k^2}{m \beta}) \, \int_0^t d\tau 
(t-\tau) \, Z(\tau)] \, ,
\end{equation}

\noindent where $Z(t)$ stands for  
the normalized velocity autocorrelation 
function (VACF), i.e., $Z (t) = \langle \vec{v}_1(t) \vec{v}_1(0) \rangle
/ \langle v_1^2 \rangle $, 
of a tagged particle in the fluid. 
This approximation gives correct results in the limits of both 
small and large wavevectors
and it also recovers the correct short time behaviour. Therefore, 
the evaluation of $F_s(k, t)$ 
is closely connected with the dynamics of a tagged 
particle in the fluid, which will be calculated as described  
in the next section. 

For completeness we also mention here the 
viscoelastic approximation for $F(k,t)$.
This model, which amounts to approximate $\Gamma(k, t)$ by an 
exponentially decaying function with a single relaxation time
\cite{Copley&Lovesey,Lovesey}, 
has been used in previous studies of single particle dynamics within
the mode-coupling formalism as a simple and convenient expression 
for the intermediate scattering function. 
In this work, we will also use the viscoelastic model, with the relaxation 
time proposed by Lovesey \cite{Lovesey}, in order to compare its predictions 
with those derived from the more ellaborated self-consistent formalism
proposed above.

\subsection{Single particle dynamics.}

\label{single}

The self-intermediate scattering function probes the single particle 
dynamics 
over different scales of length, ranging from the hydrodynamic limit 
($k \to 0$) to the free particle limit ($k \to \infty$). Within the 
gaussian approximation (eq.\  \ref{FsuS}), its evaluation requires 
the previous 
determination of the VACF. This is a central magnitude within the 
single particle dynamics, whose associated transport coefficient is the 
self-diffusion coefficient, $D$ 
 
\begin{equation}
\label{DifZ}
D = \frac{1}{\beta m} \int_0^{\infty} dt \, Z(t) \, .
\end{equation}

In a similar way as we have proceeded for the 
intermediate scattering function, 
$F(k, t)$, the dynamical features related to the motion of a 
tagged particle in the fluid are studied 
by resorting to its first-order memory function, $K(t)$, 
defined as 

\begin{equation}
\label{MemZ}
\dot{Z}(t)= -\int_0^t \; dt' \; K(t - t') \; Z(t') \, .
\end{equation}


Now, $K(t)$ is  
split into binary  
and mode-coupling contributions \cite{Sjobook,Sjothself}, i.e.,  

\begin{equation}
\label{Ktot}
K(t) = K_B(t) + K_{\rm MC}(t) \, ,
\end{equation}

\noindent where the first contribution, $K_B(t)$,    
describes single uncorrelated 
binary collisions between a tagged particle and another one from 
the surrounding medium,  
whereas the second one, $K_{\rm MC}(t)$, describes the long-time 
behaviour and originates 
from long-lasting correlation effects among the collisions.   
Below, we briefly describe both contributions.

\subsubsection{The binary term.}  
At very short times $K(t)$ is well described by $K_B(t)$ only;
moreover both have the same initial 
value ($\Omega_0$, the Einstein frecuency)
and the same initial time decay ($\tau_D$), which, within the superposition
approximation for the 3-body distribution function, is given as


\begin{eqnarray}
\label{TauD}
\frac{\Omega_0^2}{{\tau_D}^2} & = &
\frac{\rho}{3 m^2} \int d \vec{r}\, 
[\nabla^{\alpha}\nabla^{\beta}\varphi(r)]\, g(r) \,
[\nabla^{\alpha}\nabla^{\beta}\varphi(r)]  \nonumber \\
& + & \frac{1}{6 \rho} \int \frac{d {\vec{k}}^{\prime}}{(2 \pi)^3} 
\gamma_d^{\alpha \beta}(k^{ \prime}) 
\ [ S(k^{ \prime}) -1 ]  
 \gamma_d^{\alpha \beta }(k^{ \prime}) \,\,\,\, . 
\end{eqnarray}

\noindent This quantity can therefore be evaluated 
in terms of the interatomic pair potential and the static 
structural functions of the liquid. A reasonable, semi-phenomenological,  
 expression for the binary term  
is provided by the following gaussian ansatz   

\begin{equation}
\label{kb}
K_B(t) = \Omega_0^2 \;  e^{-t^2/\tau_D^2}\, ,
%
\end{equation}

\noindent which allows the calculation of $K_B(t)$ from the knowledge of the 
static structural functions only. 

\subsubsection{The mode-coupling term.} 
The presence of a slowly varying part in 
$K(t)$ is essential  
for the correct description of the dynamics of
a tagged particle in a fluid \cite{Levesque}. This is taken into account by
the mode-coupling contribution, which arises from non-linear couplings of 
the velocity of the particle with other modes of the fluid. Although 
coupling to several modes should be
considered, it has been shown 
\cite{Shimojo,Gudowski,SjoJPC} that for thermodynamic conditions near the 
triple point, the most important contribution  
arises from the density-density coupling and therefore we   
write \cite{SjoJPC,GGCan}

\begin{eqnarray}
\label{Kmc}
K_{\rm MC}(t) = \frac{\rho}{24 \beta \pi^3 m} \int d\vec{k} \; & k^2 &
\, c^2(k) \left[ F_s(k,t) F(k, t) \right.  \nonumber \\
& - & \left. F_{sB}(k,t) F_B(k,t) \right]  \, . 
\end{eqnarray}

As already mentioned, Balucani {\em et al} \cite{BaluD,TorBalVer} 
have studied  
some dynamical properties of the liquid alkali metals by using, basically, 
the same coupling term as given in equation (\ref{Kmc}) with the 
$F(k, t)$ taken either from MD simulations or evaluated 
within the viscoelastic model. 
They found that 
a substantial contribution to the integral in equation (\ref {Kmc}) 
comes from the most slowly decaying density fluctuations, namely,
those in the region around the first maximum, $k_p$, of $S(k)$ 
and therefore they simplified the evaluation of equation (\ref{Kmc}) by 
considering only a weighted contribution  
of $k=k_p$ to the integral. This approach 
was also followed in another  
study of liquid lithium at several 
thermodynamic states \cite{TorBalVer}; now by studing the integrand in 
equation (\ref{Kmc}) they concluded that besides the 
$k=k_p$ contribution, another one from $k \approx 0.65 k_p$  
should also be taken into account. 
However, in a MD study for liquid sodium, Shimojo {\em et al} 
\cite{Shimojo} have suggested that the whole range
of wavevectors should be evaluated in order to quantitatively 
account for the 
mode-coupling effects on the memory functions. Therefore, in recent 
studies of liquid lithium \cite{GGCan} and liquid alkaline-earth metals 
\cite{Alemany} close to the triple point, we 
have incorporated the whole integral in equation (\ref{Kmc}) leading to 
an improved description of the single particle properties.

\subsection{Self-consistent procedure.}

\label{self}

By approximating the intermediate scattering function, $F(k, t)$, by that 
obtained within the viscoelastic 
model \cite{Copley&Lovesey}, we developped 
a self-consistent scheme \cite{GGCan} by which we studied the 
single particle dynamics, as represented by the VACF,  
its memory function, the self-diffusion coefficient, and the 
self dynamic structure factor. Now, we extend this scheme to include 
the self-consistent calculation of the collective properties, as represented 
by the intermediate scattering function, its second-order memory function 
and the dynamic structure factor.  

We start with an estimation for both the mode-coupling component of  
$K(t)$, (e.g., $K_{\rm MC}(t)=0$) and the intermediate scattering 
function, $F(k, t)$ (e.g., the viscoelastic model). 
Using the known values of 
$K_B(t)$ and equation (\ref{Ktot}), a VACF total memory function 
is obtained which, 
when taken to equation (\ref{MemZ}), gives a normalized VACF. 
Now, by using the gaussian approximation for $F_s(k,t)$, 
as given by equation (\ref{FsuS}),  
the evaluation of the integral   
in equation (\ref{Kmc}) leads to a new estimate for the 
mode-coupling component 
of the memory function. Finally,
this ``single particle'' loop is iterated until self-consistency 
is achieved between the initial
and final VACF total memory function, $K(t)$.
Now, from the previously obtained dynamical magnitudes, and using the 
known values of $\Gamma_B(k, t)$, we evaluate the 
mode-coupling term, $\Gamma_{\rm MC}(k, t)$, as given in 
equation (\ref{gammamc}), obtaining a total second-order memory function. 
The resulting values are taken to equations (\ref{LapF})- (\ref{MfF}) 
leading to a new estimate of the intermediate scattering function, $F(k, t)$. 

Now, with this new value for $F(k, t)$, we return 
back to the ``single particle'' 
loop, which is once more iterated towards self-consistency; therefrom,   
a new estimate for $F(k, t)$ is obtained and  again 
the whole procedure is repeated 
until self-consistency is achieved for both the $K(t)$ and the $F(k, t)$. 
The practical application of the above scheme has shown that reaching 
self-consistency for the ``single particle'' loop requires around five  
iterations, whereas the whole process of getting self-consistency for 
both the $K(t)$ and the $F(k,t)$ needs between six and ten iterations.  

In this paper, we apply this scheme to study liquid lithium at several 
thermodynamic states. The results obtained within this scheme will 
usually be compared with those predicted by the viscoelastic approximation, 
that is, those results obtained using the viscoelastic $F(k, t)$ 
and performing just a ``single particle'' loop.

Finally, we end up by signalling that the self-diffusion coefficient 
may be computed either via equation  
(\ref{DifZ}) or also from the VACF total memory function, as

\begin{equation}
\label{difusion}
D^{-1} = m \beta \int_0^{~\infty} dt \: K(t) \, . 
\end{equation}

\subsection{Shear viscosity.}
\label{shear}

Another interesting transport property is the shear viscosity 
coefficient, $\eta$, which can be obtained as the time integral of the 
stress autocorrelation function (SACF), $\eta(t)$. This latter magnitude 
stands for the time autocorrelation function of the non-diagonal elements 
of the stress tensor, see eqn. (\ref{etaTOT}) below. 
Moreover, $\eta(t)$ can be decomposed into three contributions, a
purely kinetic term, $\eta_{kk}(t)$, a purely potential 
term, $\eta_{pp}(t)$, and a crossed term, $\eta_{kp}(t)$. However,  
for the liquid range close to the triple point,  
the contributions to $\eta$ coming from the first and last
terms are negligible \cite{Balueta}, and therefore in the present 
calculations we just consider  
$\eta_{pp}(t) \equiv \eta(t)$, the purely
potential part of the SACF. Now, this function is again split
into two contributions, namely, a binary and a mode-coupling
component respectively, 
$\eta(t) = \eta_B(t) + \eta_{\rm MC}(t)$. Again, the binary part 
is described by means of a gaussian ansatz, i.e.,   

\begin{equation}
\eta_B(t) = G_p \; e^ {-t^2/\tau_{\eta}^2} \, , 
\end{equation}

\noindent where $G_p$, the rigidity modulus, is the initial value of both  
$\eta(t)$ and $\eta_B(t)$, and $\tau_{\eta}$ is their initial time 
decay. As shown in Ref.\  \onlinecite{GGCan}, both $G_p$ and $\tau_{\eta}$  
can be computed from the knowledge of the interatomic potential and 
static structural functions of the system. The superposition approximation 
for the three-particle distribution function is also used in the evaluation 
of $\tau_{\eta}$.   

The mode-coupling component, $\eta_{\rm MC}(t)$, takes again into 
account the
coupling with the slowly decaying modes, and for the liquid range 
studied in
this work is basically dominated by the coupling to density 
fluctuations. Its 
expression, within the mode-coupling formalism is 
given by \cite{Balubook,GGCan,Balueta},

\begin{eqnarray}
& &\eta_{\rm MC}(t) = \nonumber \\
& &\frac{1}{60 \beta \pi^2} \int dk \, k^4 \left[ \frac{S
^{\prime}(k)}{S^2(k)} \right]^2 \left[ F^2 (k,t) - 
F_B^2(k,t) \right] \, ,
\end{eqnarray}

\noindent where $F_B(k, t)$ is given by equation (\ref{FsB}) and  
$S^{\prime}(k)$ is the derivative of the static 
structure factor with respect to $k$. This  
mode-coupling integral is evaluated by using the $F(k, t)$ and $F_B(k, t)$ 
previously obtained within the self-consistent scheme.

\section{Molecular Dynamics simulations.}

\label{simul}

We have simulated liquid $^7$Li at the three thermodynamic states specified 
at the Table \ref{thermodynamic}; one state close to the triple 
point (T=470 K) and the other two at slightly higher 
temperatures (T=526 and 574 K). The simulations have been performed by
using 668 (T=470 K), 740 (T=526 K) and 735 (T=574 K) particles 
enclosed in a cubic box with periodic boundary conditions.
For the integration of the equations of motion we used Beeman's
algorithm \cite{Beeman}, with a time step of 3 fs. The properties  
have been calculated from the configurations generated during a run 
of 10$^5$ equilibrium time steps after an equilibration period of 10$^4$  
time steps. Moreover, in the case of T=470 and 574 K, we have also 
evaluated several  
$k$-dependent properties; for T=470 K we have considered 
20 different $k$-values in a   
$k$-range 
between $0.25 $ \AA$^{-1}$ and $5 $ \AA$^{-1}$, whereas for T=574 K we 
considered 17 different $k$-values in the same $k$-range.

The computation of both the normalized VACF and 
mean square displacement has been performed according to the standard 
procedures \cite{Allen}.
The VACF total memory fucntion is 
calculated by solving the equation (\ref{MemZ}) as follows. From the computed 
values of the VACF, we obtain its time derivative, i.e., the left hand side 
of equation (\ref{MemZ}). Then, we start with an initial estimation 
for $K(t)$ consisting of 50 points uniformly distributed within a range 
$0-t_{\rm max}$ (usually, 1 ps $ \le t_{\rm max} \le $ 2 ps), from 
which a $K(t)$ for all 
times is constructed by cubic splines. By evaluating the right hand side 
of equation (\ref{MemZ}), an estimation is now obtained 
for $\dot{Z}(t)$ which is 
compared with its exact value so as to obtain the mean square deviation of 
the initial estimation for $K(t)$. The software 
package MERLIN \cite{Merlin} is now used to optimize the 50 values of the 
$K(t)$ by minimizing such a mean square deviation.

The computation of $F(k,t)$ and $F_s(k,t)$ follows from its definition
\cite{npamd2}, whereas the evaluation of the first and 
second-order memory functions of $F(k, t)$ has been performed by means 
of the same approach as previously mentioned for the  
VACF total memory function. The dynamic structure factors $S(k,\omega)$ and
$S_s(k,\omega)$ were obtained by Fourier 
transformation of the corresponding intermediate
scattering functions; however, in the case of $F_s(k,t)$ 
for small $k$, it is not
possible to use this straightforward procedure, since these functions decay
very slowly, and another procedure, described in 
Ref.\  \onlinecite{npamd2}, was used to obtain $S_s(k,\omega)$. 

Finally, the shear viscosity coefficient, $\eta$, has been obtained by 
the time integral of the SACF, from the following 
Green-Kubo-like relation

\begin{equation}
\label{etaTOT}
\eta = \frac{\beta}{3 V} \int_0^{\infty} \sum_{\alpha \gamma} \; 
\langle \sigma^{\alpha \gamma}(t) \sigma^{\alpha \gamma}(0) \rangle \;  dt
\equiv \int_0^{\infty} \eta (t) \, ,
\end{equation}

\noindent where the sum is to be made 
on the circular permutation of the 
indices $\alpha \gamma$ ($xy, \; yz, \; zx$), $V$ is the volume of 
the system and  
$\sigma^{\alpha \gamma}(t)$ are the time dependent elements of the 
microscopic stress tensor, namely,

\begin{equation}
\sigma^{\alpha \gamma}(t) = m \sum_i^N v_i^{\alpha}(t) v_i^{\gamma}(t) + 
\sum_{1=i<j}^N r_{ij}^{\alpha}(t) F_{ij}^{\gamma}(t) \, ,
\end{equation} 

\noindent which is computed in terms of the velocities $v_i^{\alpha}(t)$,  
relative forces $F_{ij}^{\gamma}(t)$, and relative 
distances $r^{\alpha}_{ij}(t)$ between the particles.

\section{Results.}

\label{litio}

We have applied the preceeding theoretical formalism to study the dynamical
properties of liquid $^7$Li at three thermodynamic states close to 
the triple point, which correspond to those of the neutron scattering 
experiments (see Table I).
The input data required by the theory are 
both the interatomic pair potential and its derivatives as well as 
the liquid static structural properties, $g(r)$, $S(k)$, $c(k)$. 
We have recently proposed
a new interatomic pair potential for liquid lithium, based on 
the Neutral Pseudo Atom
(NPA) method \cite{npavmhnc}, which,
when used in conjunction with the Variational
Modified Hypernetted Chain approximation (VMHNC) theory of liquids
\cite{Ros86,PRA}, has
proved to be highly accurate for the calculation of the liquid static
structure and the thermodynamic properties. Moreover, MD
simulations using this potential \cite{npamd} have shown that the dynamical
properties are also well reproduced when compared with the available
experimental data. We will therefore use both the NPA 
interatomic pair potential and the
corresponding liquid static structure results, obtained 
within the VMHNC theory, as
the input data required for the calculation of the dynamical properties. We
want to stress here that the only input parameters needed for 
the calculation of
the interatomic pair potential and the liquid static structure are 
the atomic number of
the system and its thermodynamic state (number density and temperature).

\subsection{Intermediate scattering function.} 

The intermediate scattering function, $F(k, t)$, embodies the information 
concerning the collective dynamics of density fluctuations over both the 
length and time scales. Moreover, it 
is a basic ingredient for the evaluation of both collective and 
single particle properties; therefore it is important to achieve an accurate 
description of this magnitude.

\begin{figure}
\begin{center}
\mbox{\psfig{file=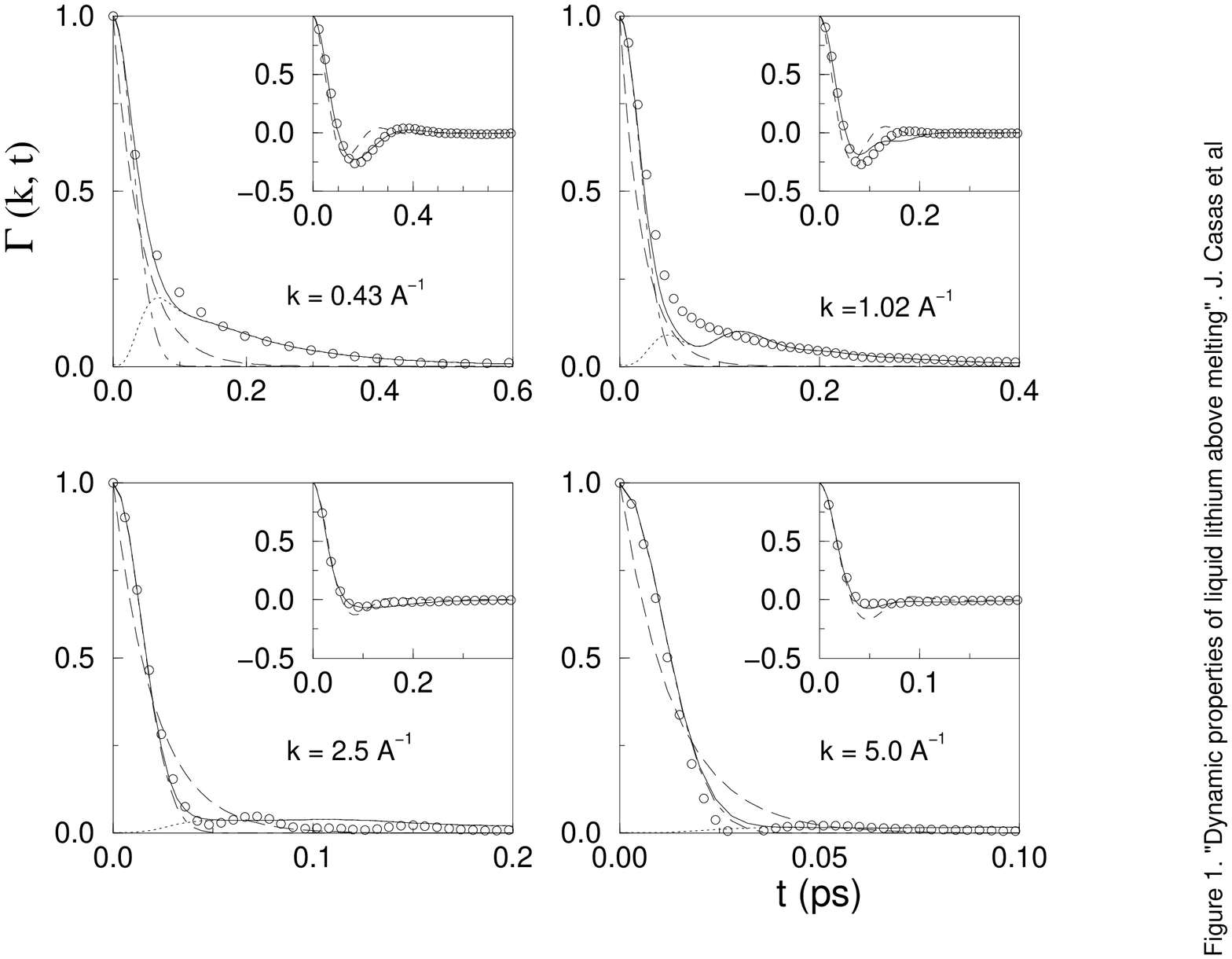,angle=-90,width=85mm}}
\end{center}
\caption{Normalized second-order memory function, $\Gamma (k, t)$, of 
the intermediate scattering functions, $F(k, t)$, at  
several $k$-values, for liquid 
lithium at T = 470 K. Open circles: MD results. 
Continuous line: present theory. 
Dash-dotted line: binary part, $\Gamma_B(k, t)$. 
Dotted line: mode-coupling part, $\Gamma_{\rm MC}(k, t)$. 
Dashed line: viscoelastic model. The inset shows the normalized 
first-order memory 
function as obtained by MD (open circles), the viscoelastic 
model (dashed line) and the present theory (continuous line).}
\label{MEMFKT470}
\end{figure}

\begin{figure}
\begin{center}
\mbox{\psfig{file=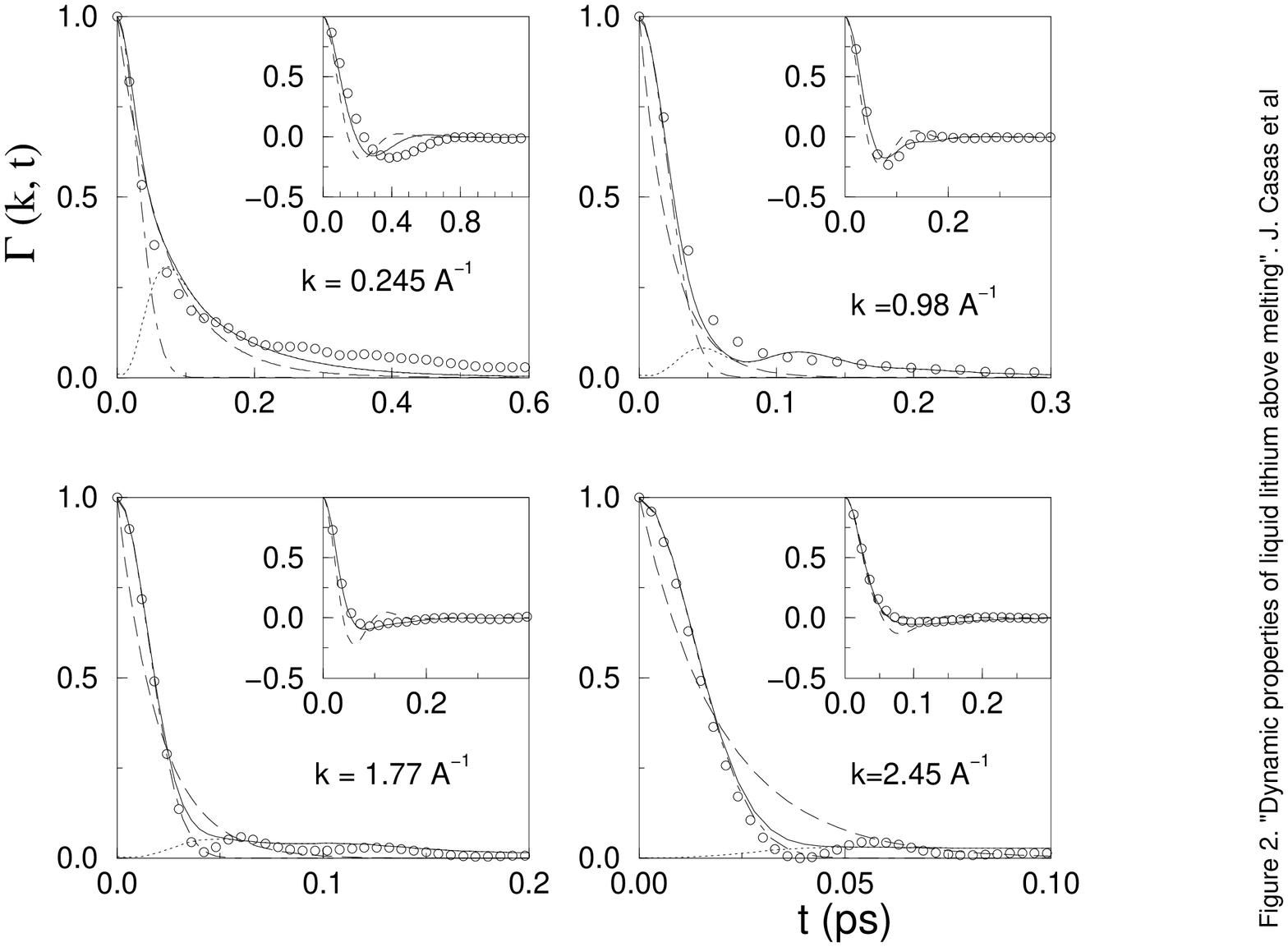,angle=-90,width=85mm}}
\end{center}
\caption{Same as the previous figure but for T = 574 K. } 
\label{MEMFKT574}
\end{figure}

First, in figures \ref{MEMFKT470}-\ref{MEMFKT574} we show, 
for T=470 K and 574 K at several $k$ values, the 
MD results obtained for the 
first and second-order memory functions of the $F(k, t)$. 
The MD results for the second-order memory function, $\Gamma_{\rm MD}(k, t)$,  
show that for $k \leq k_p$ it remains 
positive at all times, whereas for greater $k$'s it can 
take negative values. 
It displays a rapidly decaying part at small times and 
a long-time tail which becomes longer for the smaller $k$-values, and 
it exhibits  
an oscillatory behaviour for $k \geq 1.3 \approx k_p/2$. On the other hand, 
the corresponding MD results for the first-order 
memory function, exhibit a negative minimum at all the $k$-values 
considered, although its absolute value decreases for increasing $k$'s and 
becomes rather small for  
$k \approx k_p$; moreover it exhibits weak oscillations for the 
small $k$-values. In fact, very similar qualitative features were 
also obtained by 
Shimojo {\it et al} \cite{Shimojo} in their MD study for liquid Na
near the triple point. 

Figures \ref{MEMFKT470}-\ref{MEMFKT574} also show 
the theoretical $\Gamma (k, t)$ obtained 
within the iterative scheme described in section \ref{self}. As 
already mentioned, a number of around eight  
iterations were enough to achieve self-consistency for this magnitude. 
First, we note that the theoretical $\Gamma (k, t)$ 
reproduce, qualitatively, the corresponding MD results. 
In particular, the short time behaviour, which is dominated by the 
binary component, is very well described. This is not unexpected, 
as the present theoretical 
formalism imposes the exact initial values of both $\Gamma (k, t)$ and 
its second derivative through equations (\ref{Gambin}) and 
(\ref{Tauele}) respectively.  
Moreover, it is important to note that the overall amplitude of the decaying 
tail of $\Gamma(k,t)$ is also, in general, well described by the present 
formalism. The main discrepancies with the MD results are the amplitudes of
the oscillations around the decaying tail, which occur 
for intermediate times where the mode-coupling 
component, $\Gamma_{\rm MC}(k, t)$, dominates.  We consider that  
the main reason relies on the 
expression taken for the $\Gamma_{\rm MC}(k, t)$, see 
equation (\ref{gammamc}), 
where we have only included the density-density coupling term. 
Although with this term only, it is possible  
to reproduce the main features of the $\Gamma_{\rm MD}(k, t)$, further 
improvements would in principle require the inclusion of other terms, 
especially those related to the density-currents couplings. 
Furthermore, given the self-consistent nature of the present theoretical 
framework, the approximations made in the description of other magnitudes, 
such as $F_s(k, t)$ and $Z(t)$, will also exert some influence and 
therefore, a more rigorous treatment of these magnitudes would improve the 
theoretical results obtained for the $\Gamma (k, t)$. 

On the other hand, we note that the gaussian ansatz 
chosen for the description 
of the binary part, leads to theoretical results in reasonable 
agreement with the simulation results. We have verified that the choice of 
the squared hyperbolic secant in equation (\ref{gammaB}) leads to a 
worse description of the short time behaviour of the 
$\Gamma (k, t)$, with the binary term becoming 
clearly wider than the MD results for the large $k$ values. 

In figures \ref{MEMFKT470}-\ref{MEMFKT574} we have also included, for 
comparison, the results of the viscoelastic 
model \cite{Copley&Lovesey} for the second-order memory 
function, $\Gamma_{\rm visc}(k, t)$.  
In this model, this function is represented by 
a simple exponential decaying function where the only exact value is the 
initial value; this  explains the strong differences found between the 
$\Gamma_{\rm visc}(k, t)$ and the corresponding MD results.
However, it is interesting to observe that for small $k$-values the 
viscoelastic model underestimates the MD data at all times, whereas around 
the first peak and for higher wavevectors there is a compensation between 
the short time behaviour, where the viscoelastic model is too low, and the 
intermediate times, where it is too high when compared with the MD results.

\begin{figure}
\begin{center}
\mbox{\psfig{file=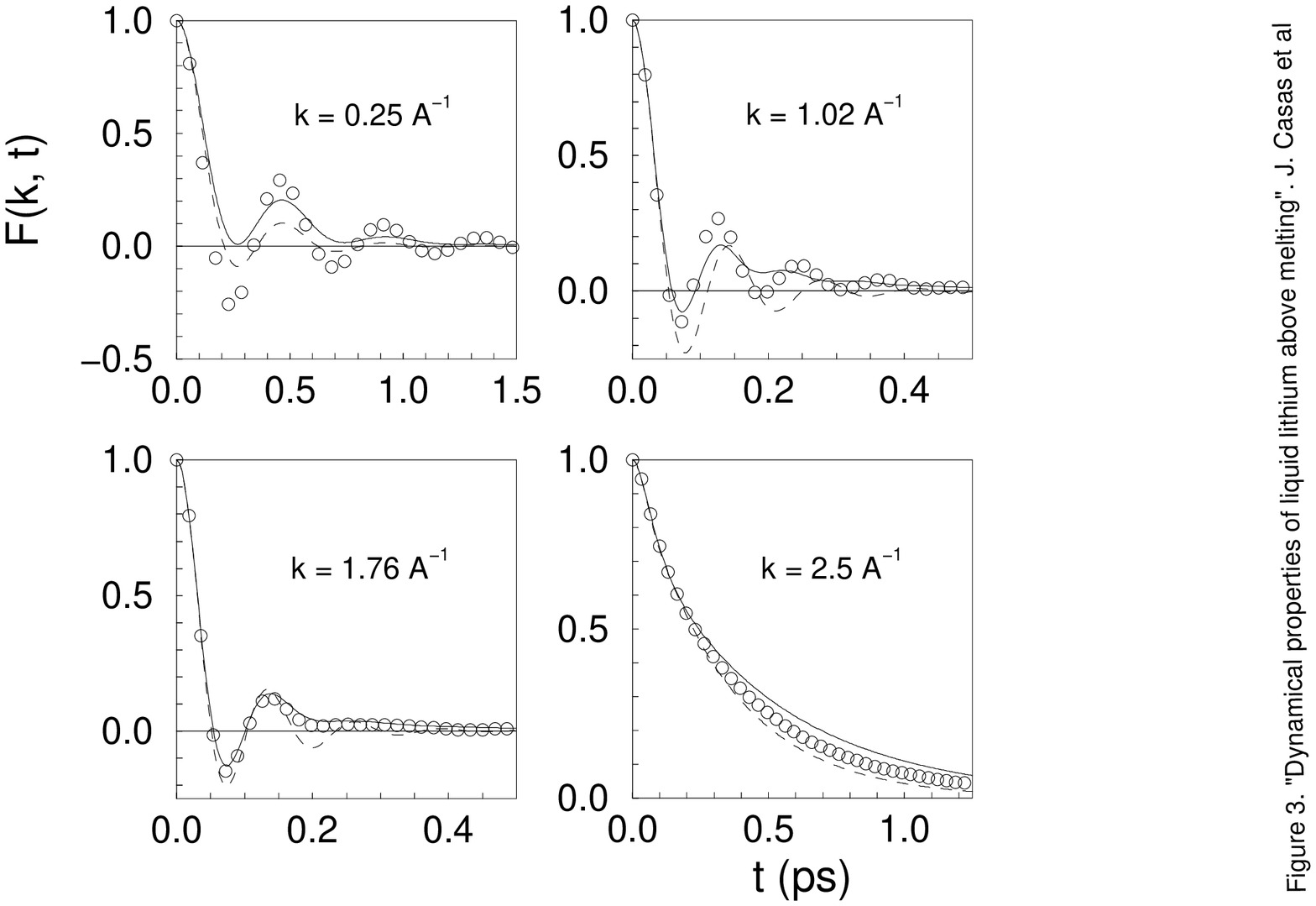,angle=-90,width=95mm}}
\end{center}
\caption{Normalized intermediate scattering functions, $F(k, t)$, at  
several $k$-values, for liquid 
lithium at T = 470 K. Open circles: MD results. 
Continuous line: present theory. 
Dashed line: viscoelastic model.}
\label{FKT470}
\end{figure}

\begin{figure}
\begin{center}
\mbox{\psfig{file=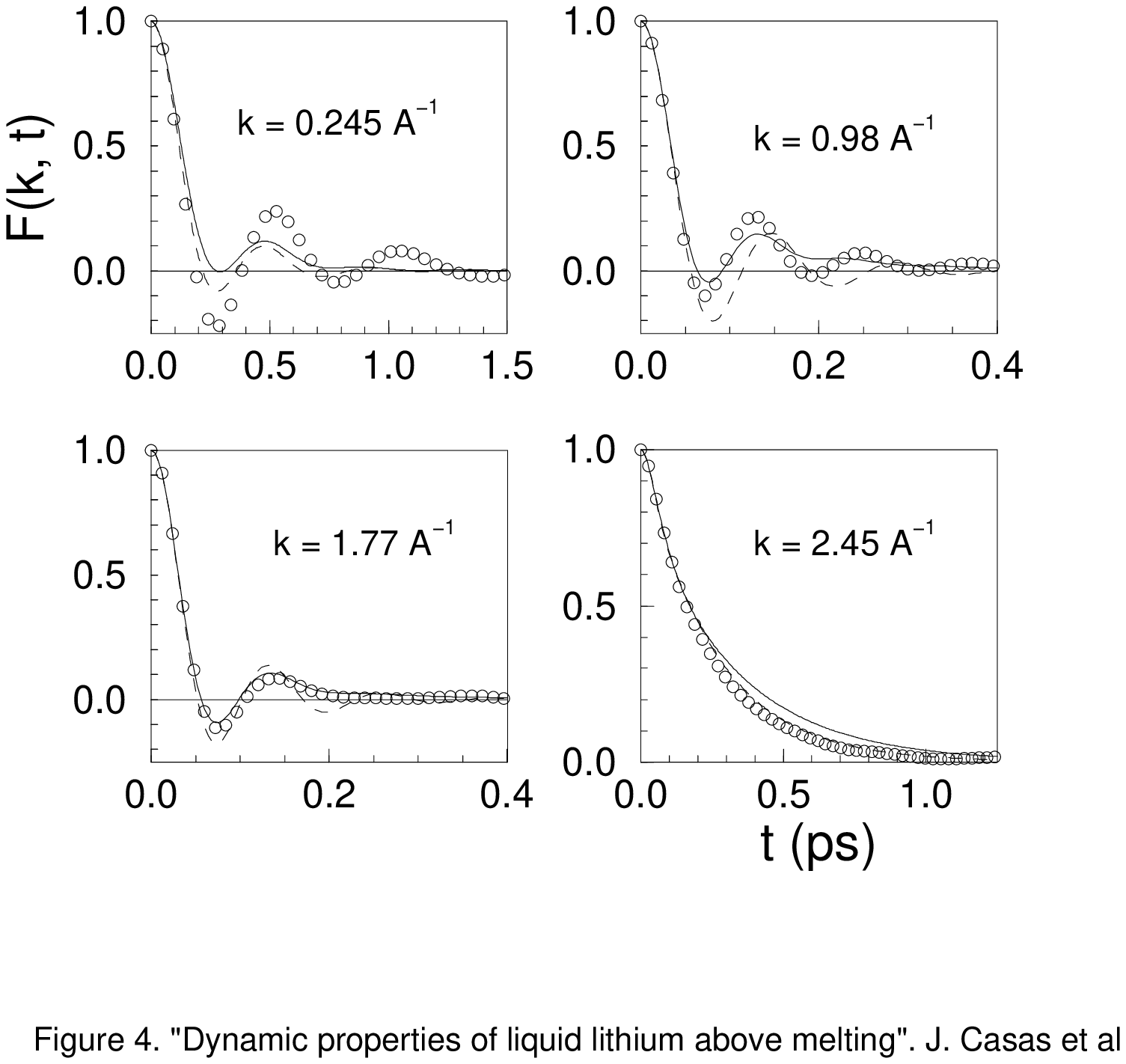,angle=-90,width=95mm}}
\end{center}
\caption{Same as the previous figure but for T = 574 K. } 
\label{FKT574}
\end{figure}

Figures \ref{FKT470}-\ref{FKT574} show, for T=470 K and 574 K at several 
$k$ values, the corresponding MD results for the 
intermediate scattering functions, $F_{\rm MD}(k, t)$. 
It is observed that the 
$F_{\rm MD}(k, t)$ exhibit an oscillatory behaviour which persists 
until $k \approx 2$ \AA$^{-1}$, with the amplitude of the oscillations
being stronger for the smaller $k$-values. These oscillations   
do not take place around zero, but instead the behaviour 
of $F_{\rm MD}(k, t)$ 
after the first minimum looks like an oscillatory one superimposed on a 
monotonically decreasing tail. 

These characteristics are qualitatively reproduced by the  
theoretical $F(k, t)$ obtained within the present formalism.
In particular, the amplitude of the decaying tail is rather well
reproduced, while the main discrepancies are located for the small 
$k$-values where the amplitude of the oscillations is underestimated. 
This is an important improvement over the results obtained within the
viscoelastic model \cite{Copley&Lovesey}, since the oscillations of
$F_{\rm visc}(k, t)$ take place basically around zero. This has important
consequences on the single-particle dynamical properties at intermediate
times, as will be shown in the next section.
The deficiencies of $F_{\rm visc}(k,t)$ are, however, not so marked as those 
previously seen for the corresponding second-order memory function. 
This improvement is explained because the viscoelastic 
model incorporates the 
exact initial values of $F(k, t)$, and its second and fourth derivatives. 
Comparison with the MD results shows that   
for the small $k$-values, i.e., $k \approx 0.25$ \AA$^{-1}$, 
the oscillations in 
the $F_{\rm visc}(k, t)$ are strongly damped, 
for the intermediate $k$-values, 
i.e., $k \approx 1$ \AA$^{-1}$ and $k \approx 1.7$ \AA$^{-1}$, the 
oscillations are stronger and they persist for longer times, 
whereas for those $k$-values around $k_p$ the results obtained by 
the viscoelastic model show a good agreement with the MD results. 
This agreement for the regions around the main peak can be understood
in terms of the compensation between the short-time and intermediate-time
deficiencies of the viscoelastic second order memory function.
By taking into account the suggestions of 
Balucani {\it et al} \cite {BaluD} already mentioned in section \ref{single},  
this good description of $F(k,t)$ provided by the viscoelastic model 
in the region of the main peak, explains why those single particle dynamics 
calculations \cite{Alemany,BaluD,GGCan,TorBalVer} based on
the viscoelastic model 
lead to results in reasonable agreement with the corresponding 
MD and experimental results.  

%
%

We stress that, to our knowledge, this is the first theoretical study 
where the intermediate scattering function, $F(k, t)$, has been evaluated 
directly from its memory function/mode-coupling formalism, without 
resorting to any parameters or fitting to any assumed shape.

\subsection{Self diffusion.}

\begin{figure}
\begin{center}
\mbox{\psfig{file=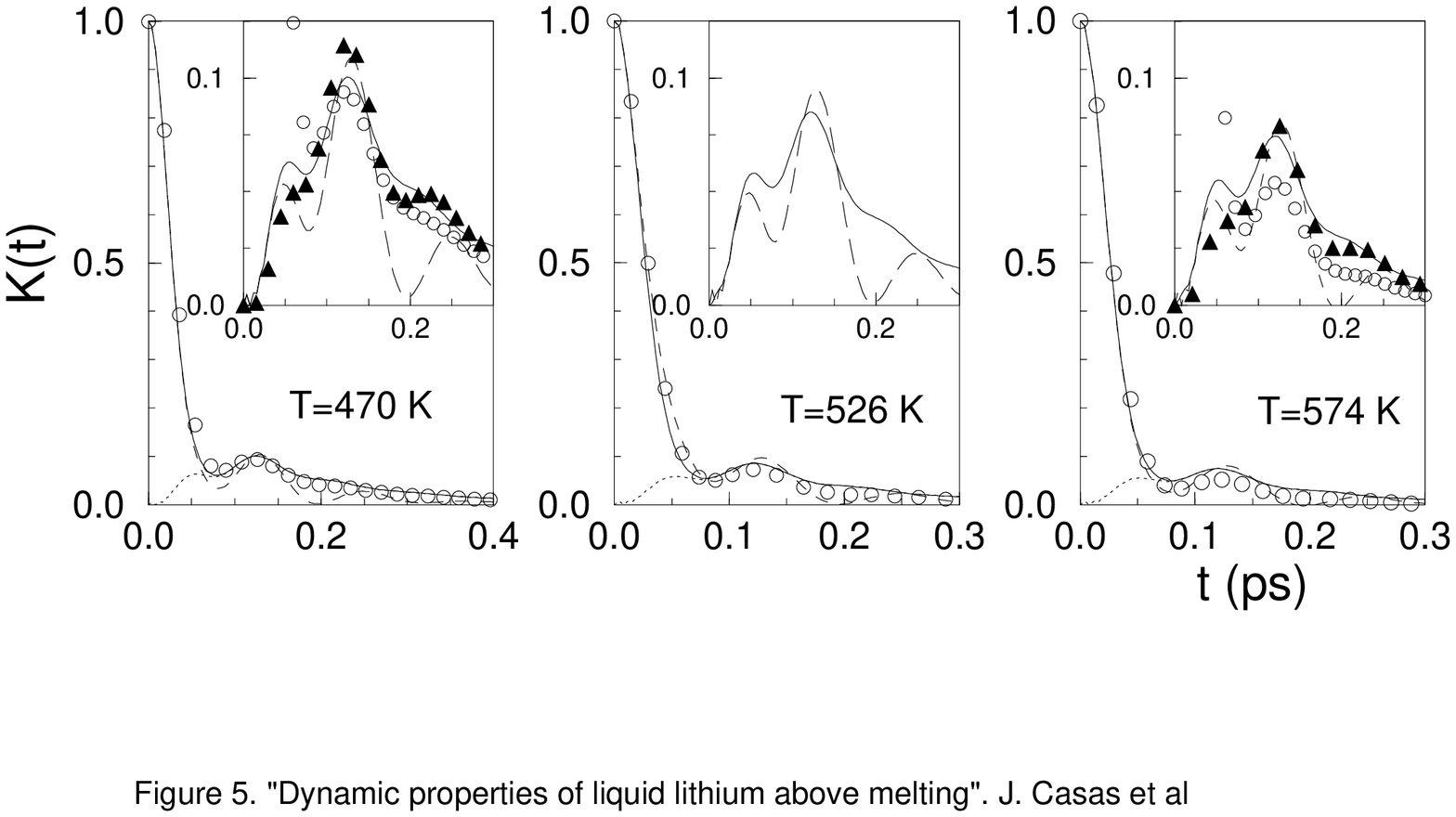,angle=-90,width=85mm}}
\end{center}
\caption{Normalized VACF total memory functions for liquid Li at three 
temperatures.  
Open circles: MD results. Continuous line: theoretical results. 
Dotted line: 
theoretical results for the mode-coupling contribution. Dashed line: 
viscoelastic approximation. The insets provide a closer comparison 
among the mode-coupling contributions from the present 
theory, $K_{\rm MC}(t)$, 
(continuous line), 
and from the viscoelastic model, $K_{\rm MC}^{\rm visc}(t)$, (dashed line). 
The insets for T=470 K and 574 K also include the MD results for the 
mode-coupling contribution $K_{\rm MC}^{\rm (MD)}$ 
(filled triangles) evaluated by  
using equation (\protect\ref{Kmc}). } 
\label{FKtot}
\end{figure}

Another magnitude obtained within the present self-consistent scheme is 
the VACF total memory function, and in figure \ref{FKtot} we 
show, for the three thermodynamic states studied in this work, 
the obtained theoretical total memory function, $K(t)$, along with 
the mode-coupling component. 
Whereas the binary part dominates the behaviour of $K(t)$ for 
$t \leq 0.06$ ps, for longer times the mode-coupling part completely 
determines the shape of $K(t)$. 
For comparison,  we have also plotted the total memory function obtained from 
the viscoelastic approximation, $K_{\rm visc}(t)$, as well as the MD memory 
function, $K_{\rm MD}(t)$, obtained when the MD normalized VACF is
used as input data in equation (\ref{MemZ}). 
First, we note that for T=470 K the present 
theoretical $K(t)$ shows a rather good agreement with the MD results, 
especially at 
the longer times where the wiggles and the tail of the 
memory function are well reproduced. This represents a significant      
improvement over previous 
results \cite{GGCan}, where the viscoelastic approximation was used and, as 
shown in figure \ref{FKtot}, 
the long-lasting tail was not qualitatively 
reproduced by the $K_{\rm visc}(t)$. 
In fact, the improvement of the present 
theoretical approach over the viscoelastic approximation can be traced 
back to the different intermediate scattering functions, $F(k, t)$,  
obtained within both theoretical schemes. By analizing which wavevectors 
are really relevant in the integral appearing in 
equation (\ref{Kmc}) it is   
found that, for any time, the dominant contribution comes from 
two sets of $k$-values centered around $k_1 \approx 1.7 $ \AA$^{-1}$ 
and $k_2 \approx k_p$, where $k_p$ ($ \approx 2.5 $ \AA$^{-1})$ is 
the position of the 
main peak of the $S(k)$. By comparing, for
these two sets of $k$-values, the different terms appearing in the 
integrand with its MD counterparts we find that the main discrepancies
come from the intermediate scattering function, $F(k, t)$ at 
$k \approx k_1$. As shown in figure \ref{FKT470}, 
both the present theoretical and the MD-based $F(k \approx k_1, t)$ show 
an oscillatory behaviour for $t \leq 0.20$ ps and 
then decrease towards zero taking positive values, whereas 
$F_{\rm visc}(k \approx k_1, t)$ keeps an oscillatory behaviour around zero 
for longer times and this is the main reason for the discrepancies  
among the corresponding memory functions observed in figure \ref{FKtot}
for $t \geq 0.20$ ps. In particular, the   
deep minimum exhibited by the $K_{\rm visc}(t)$ at $t \approx 0.2$ ps 
is connected to the fact that the $F_{\rm visc}(k \approx k_1, t)$
shows at around $t \approx 0.2$ ps a negative oscillation.  

To examine more closely the adequacy of the mode-coupling expression used in 
equation (\ref{Kmc}), we have also evaluated this expression, for T=470 K and 
574 K, by using the 
MD results for both $F(k, t)$ and $F_s(k, t)$. The obtained results, 
denoted as $K_{\rm MC}^{\rm (MD)}(t)$, are shown in the insets 
of figure \ref{FKtot}, where we have also included the corresponding 
$K_{\rm MD}(t)$. 
For $t \leq 0.08$ ps, a direct comparison between them is not possible 
due to the contribution of the binary term to the total memory function. 
For $t \ge 0.08$ ps, where the binary part is almost cero, we obtain that 
in the case of T=470 K, the 
$K_{\rm MC}^{\rm (MD)}(t)$ follows the behaviour of $K_{\rm MD}(t)$ although 
it is slightly higher; this is in qualitative agreement with the 
suggestion that, near the triple point, the effect of the other 
coupling terms not included in equation (\ref{Kmc}), gives rise 
to a small, although negative, contribution to the total memory function. 
On the other hand, the comparison of the 
$K_{\rm MC}^{\rm (MD)}(t)$ with both 
$K_{\rm MC}(t)$ and 
$K_{\rm MC}^{\rm visc}(t)$ 
reveals that the present theoretical framework 
leads to an important improvement over the results derived 
from the viscoelastic approximation, especially in the tail region. 


For T=574 K we obtain that the overestimation of 
$K_{\rm MC}^{\rm (MD)}(t)$ with respect to $K_{\rm MD}(t)$ increases, 
so this behavior
suggests that the expression used for $K_{\rm MC}(t)$ in equation (\ref{Kmc}) 
becomes less adequate. 
Although the expression seems acceptable near the triple point, it becomes
oversimplified at increasing temperatures and it seems necessary 
to take into account the coupling to  other modes (i.e., the modes
associated with the currents).  
In fact, as shown by Shimojo {\it el al} \cite{Shimojo} in their MD study 
for liquid Na, the combined effect 
of the coupling terms associated with the currents, gives rise to a small and 
negative contribution to the $K_{\rm MC}(t)$ which, if taken into account, 
would lead to a better agreement between the 
$K_{\rm MC}^{\rm (MD)}(t)$ and $K_{\rm MD}(t)$. 
On the other hand, it is also observed that the theoretical  
$K_{\rm MC}(t)$ obtained through the present self-consistent scheme
follows nicely the $K_{\rm MC}^{\rm (MD)}(t)$, whereas the deficiencies in
the viscoelastic $K_{\rm MC}^{\rm visc}(t)$ become more 
marked as the temperature 
is increased, with its minimum at $t \approx 0.2$ ps 
becoming deeper and taking on negative values. 


The results obtained for the self-diffusion coefficients, $D$,  
are shown  
in Table \ref{diffu} for the three thermodynamic states considered in this 
work. Specifically, the MD results have been deduced from the slope of 
the corresponding mean square displacement whereas the theoretical 
ones where obtained from equation (\ref{difusion}). First, we note that the 
MD results compare rather well with the different experimental data 
(INS data \cite{Jong1,Seldmeier},  
tracer data \cite{Lowen}) for the three temperatures.  
The theoretical results show good agreement with the MD ones for  
T=470 K whereas those for T=526 K and T=574 K are slightly underestimated.  
In fact, according to the relation between $D$ and 
the VACF total memory function, 
see equation (\ref{difusion}), this underestimation of $D$ is a   
consequence of the above mentioned overestimation of the 
mode-coupling component of $K(t)$ when the temperature is increased. 

In Table \ref{diffu} we have also included 
the values for $D$ obtained within the 
viscoelastic approximation. According to the shortcomings already mentioned 
for the corresponding memory functions, $K_{\rm visc}(t)$, obtained within  
this approximation, when equation (\ref{difusion}) is applied to compute the 
corresponding $D$, it leads to values which overestimate both the  
MD and experimental results, excepting that for T=574 K. 
Nevertheless, the reasonable agreement achieved at the 
higher temperature is, basically due to the cancellation between the 
overestimation and the underestimation 
of the $K_{\rm visc}(t)$ at $t \approx 0.13$ ps. and 
$t \approx 0.20$ ps. respectively. 

Finally, we end up by noting that the gaussian ansatz chosen for the binary 
part of the $K(t)$, provides an accurate description of its short 
time behaviour. We have performed calculations by using the squared hyperbolic 
secant in equation (\ref{kb}), and the obtained, selfconsistent,  
results overestimate both the binary and mode-coupling parts of the total 
memory function; in fact it predicts, for T=470 K, a 
self-diffusion coefficient, $D$=0.55 \AA$^2$/ps which 
is significantly smaller than the corresponding MD result.  

\subsection{Dynamic structure factors.}

In this work, both $K(t)$ and $F(k, t)$ are 
obtained within a self-consistent scheme, 
whereas the self intermediate scattering function, $F_s(k, t)$, 
is computed within the 
gaussian approximation, see equation (\ref{FsuS}), by using 
the VACF deduced from the self-consistent process. Now, by 
Fourier tranforming $F(k, t)$ and $F_s(k, t)$ we get 
$S(k, \omega)$ and $S_s(k, \omega)$ respectively, and by 
equation (\ref{Stot}), the $S_{\rm tot}(k, \omega)$ is finally obtained. 

\begin{figure}
\begin{center}
\mbox{\psfig{file=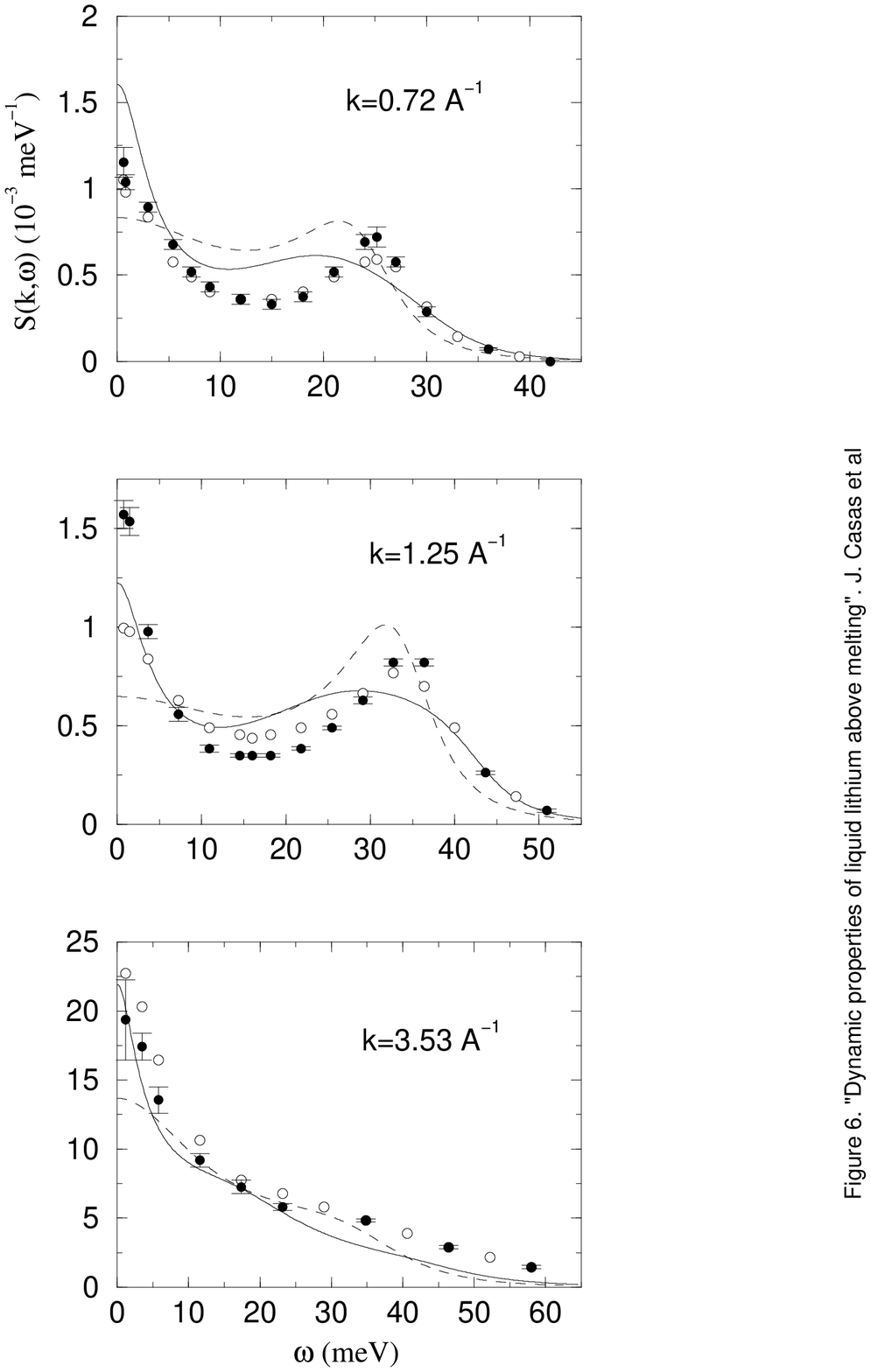,angle=0,width=85mm}}
\end{center}
\caption{Dynamic structure factors of liquid Li at T=470 K.
Continuous line: theoretical results.
Dashed line: viscoelastic approximation.
Open circles: MD results.
Full circles with error bars: experimental
IXS data 
\protect\cite{Sinn}. }
at T=488 K
\label{figSkw}
\end{figure}

In figure \ref{figSkw} we show the dynamic structure factors $S(k,\omega)$
obtained from the present formalism, along with the MD results \cite{npamd2}, 
for T=470 K, and the experimental IXS data \cite{Sinn}, measured at T=488 K, 
for three wave-vectors.
The viscoelastic $S_{\rm visc}(k,\omega)$ are also included for comparison.
The good agreement between experiment and MD has already been remarked
\cite{Sinn}, and supports the adequacy of the potential used in order to
describe the effective interaction among the ions in the liquid.
It can be observed that the overall shape of $S(k,\omega)$ is qualitatively
reproduced by the present theoretical approach, in contrast with the
viscoelastic model, which gives correctly the peak positions, but fails to
describe the overall $\omega$-dependence of the dynamic structure factor.
The behavior of $S(k,\omega)$ is of course a consequence of the time
dependence of the intermediate scattering functions, $F(k,t)$. 
For instance, the fact that the amplitude of the 
decaying tail of $F(k,t)$ is well reproduced implies that for small
$\omega$ the structure factor will increase significantly, as observed in
the figure. On the other hand, the deficiencies in the description of the
oscillations of $F(k,t)$ around the decaying tail are reflected in the 
less accurate representation of the Brillouin peak of $S(k,\omega)$.

\begin{figure}
\begin{center}
\mbox{\psfig{file=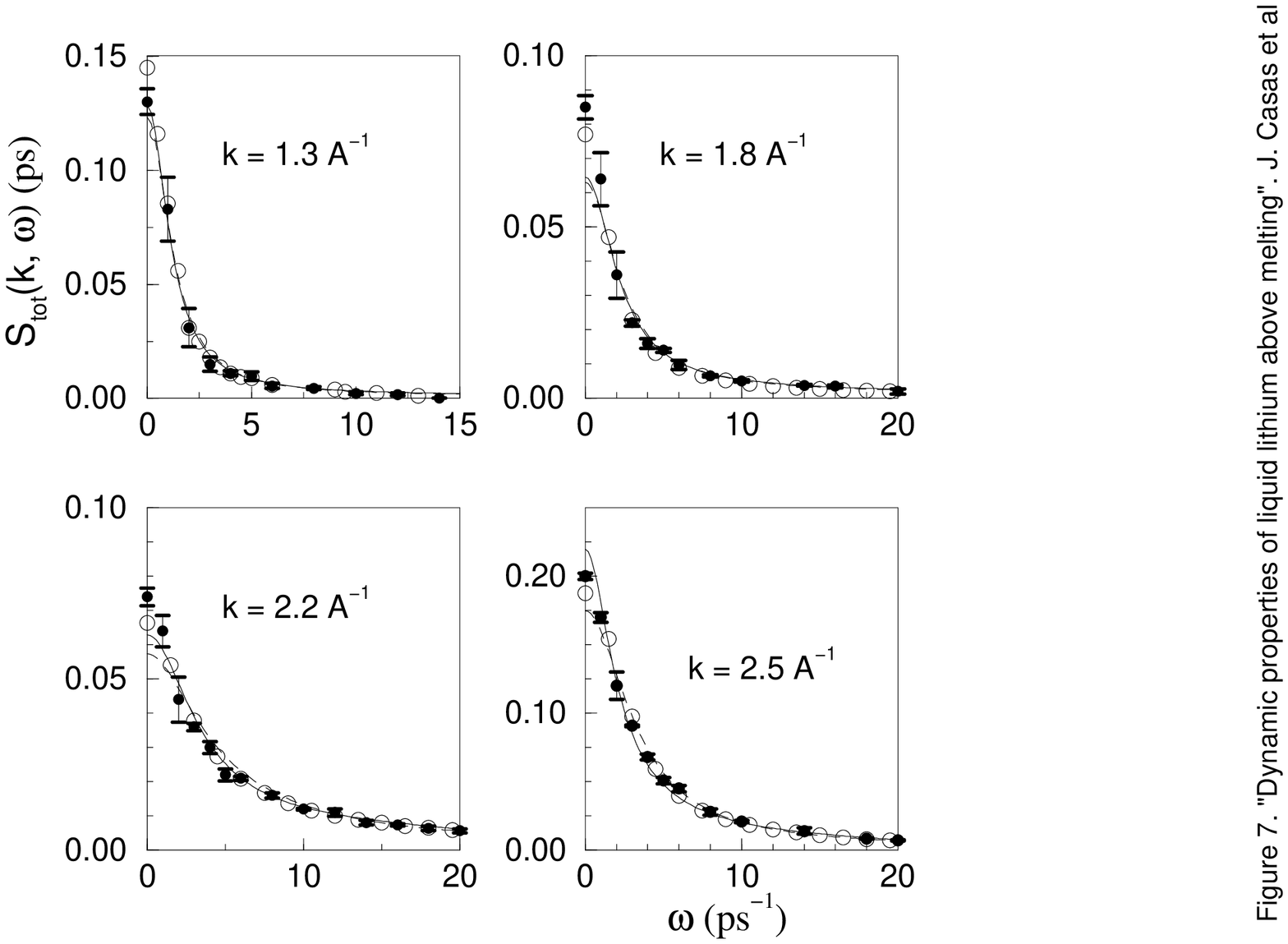,angle=-90,width=95mm}}
\end{center}
\caption{Total dynamic structure factors at several $k$-values 
corresponding to liquid Li at T= 470 K. Continuous line: theoretical results. 
Dashed line: viscoelastic approximation. Open circles: MD results. 
Full circles with error bars: 
experimental INS data \protect\cite{Jong1,Jong3}. } 
\label{figureST4}
\end{figure}

\begin{figure}
\begin{center}
\mbox{\psfig{file=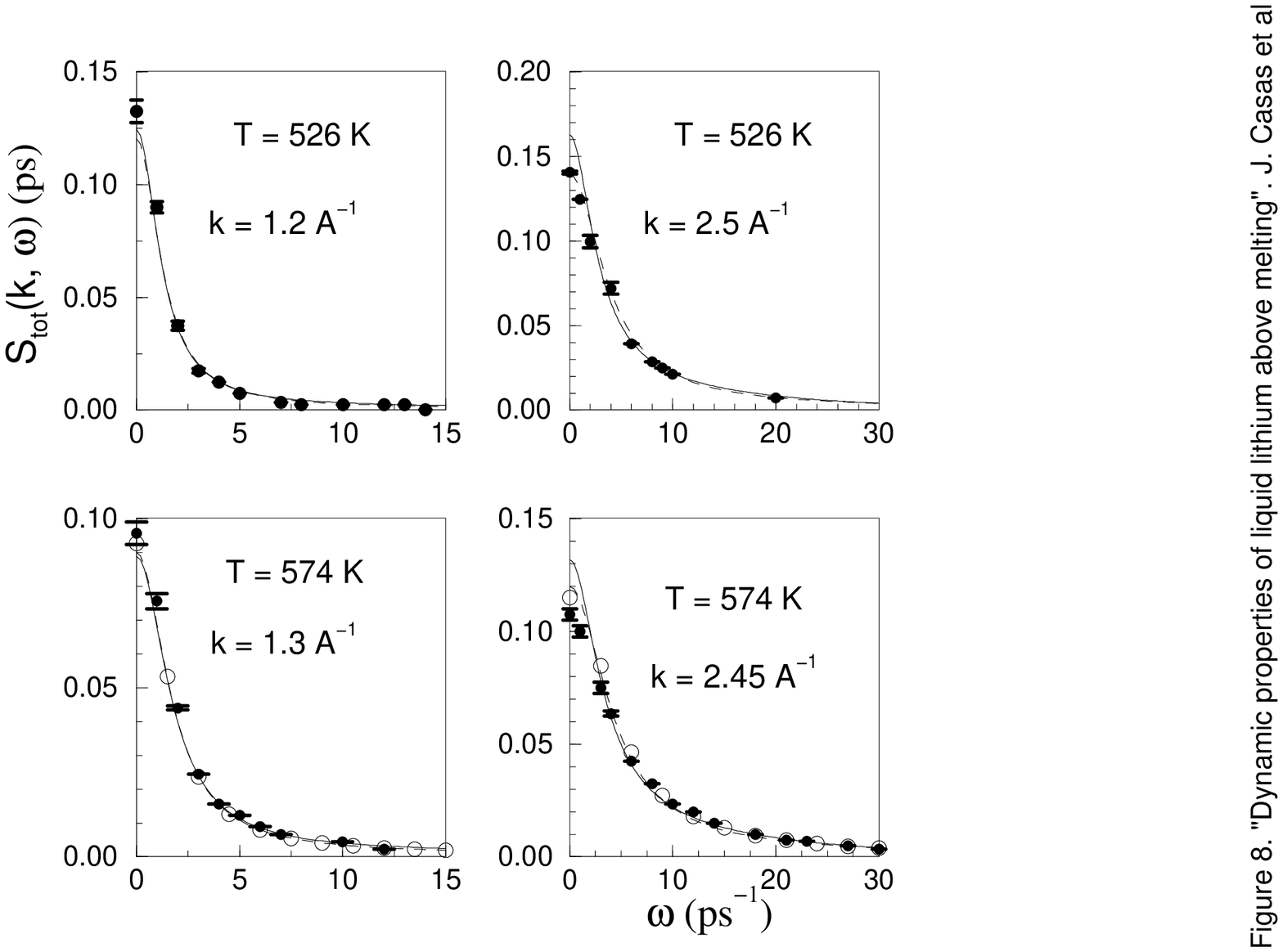,angle=-90,width=95mm}}
\end{center}
\caption{As the previous figure, but for Li at T=526 K and 574 K.} 
\label{figureST5}
\end{figure}

The theoretical results for 
$S_{\rm tot}(k, \omega)$ as a function of $\omega$, at some fixed 
$k$-values, for liquid lithium at the three temperatures, are shown in 
figures \ref{figureST4}-\ref{figureST5}, where they 
are compared with both the corresponding 
MD and experimental data \cite{Jong1,Jong2,Jong3}.  
First, it is observed that the MD results show a good agreement 
with the experimental data, except for some discrepancies appearing 
at the frequency region close to 
$\omega \approx 0$.   
Although these discrepancies could be ascribed to the  
NPA interatomic pair potential used in the simulations, this may not be 
the ultimate reason;  
in fact similar problems have also been encountered by 
Torcini {\it et al} \cite{TorBalVer} in a similar MD study carried out 
with the interatomic pair potentials proposed by    
Price {\it et al} \cite{PST}. Moreover, it must be reminded that an accurate 
experimental determination of the spectral features near 
$\omega = 0$ is not a trivial task and, therefore, the above mentioned 
discrepancies between MD and experiment must be taken with some caution. 

On the other hand, the comparison MD/theory shows that the present theory 
accounts rather well for the simulation results, except in the frecuency 
region close to $\omega = 0$.   
However, this problem has to be discussed in terms of both 
$S_s(k, \omega)$ and $S(k, \omega)$ whose combination, according to 
equation (\ref{Stot}), gives rise to the 
$S_{\rm tot}(k, \omega)$. It is observed that for $k < k_p$, the main 
contribution to the $S_{\rm tot}(k, \omega)$ comes from the $S_s(k, \omega)$,  
for $k \approx k_p$ it is the $S(k, \omega)$ the one which dominates the 
$S_{\rm tot}(k, \omega)$ whereas for $k > k_p$ both 
contributions to the $S_{\rm tot}(k, \omega)$ are 
rather similar. 
Now, when the theoretical $S_s(k, \omega)$ are compared with their MD 
counterparts, it is found that the theory can reproduce the 
MD results, except at the region close to $\omega = 0$ where the theoretical 
values are always systematically lower; in fact, this is a well known 
deficiency of the gaussian approximation used to 
calculate the self-intermediate scattering function. 
On the other hand, the theoretical $S(k, \omega)$ somewhat overestimates
the MD values near $\omega=0$, and therefore, this explains  
why the theoretical $S_{\rm tot}(k, \omega \to 0)$ are sistematically smaller 
than their MD counterparts except for those $k$-values around $k_p$.    

In the figures \ref{figureST4}-\ref{figureST5} we have also 
included the theoretical 
results obtained within the simple viscoelastic approximation. 
It is observed that 
the agreement with the MD results is very similar to that obtained with the 
present theory, although now the values for $\omega \to 0$ tend to be slightly 
smaller. This is again, basically, due to the corresponding $S_s(k, \omega)$ 
which when calculated by the gaussian approximation incorporates a 
VACF which gives a bigger diffusion coefficient. This implies that 
$F_s(k, t)$ decreases faster and leads to a smaller $S_s(k, \omega \to 0)$. 
Moreover, the values of $S_{\rm visc}(k,\omega \to 0)$ are also too low, which
reinforces the tendency to produce smaller $S_{\rm tot}(k,\omega \to 0)$

\subsection{Shear viscosity.}

\begin{figure}
\begin{center}
\mbox{\psfig{file=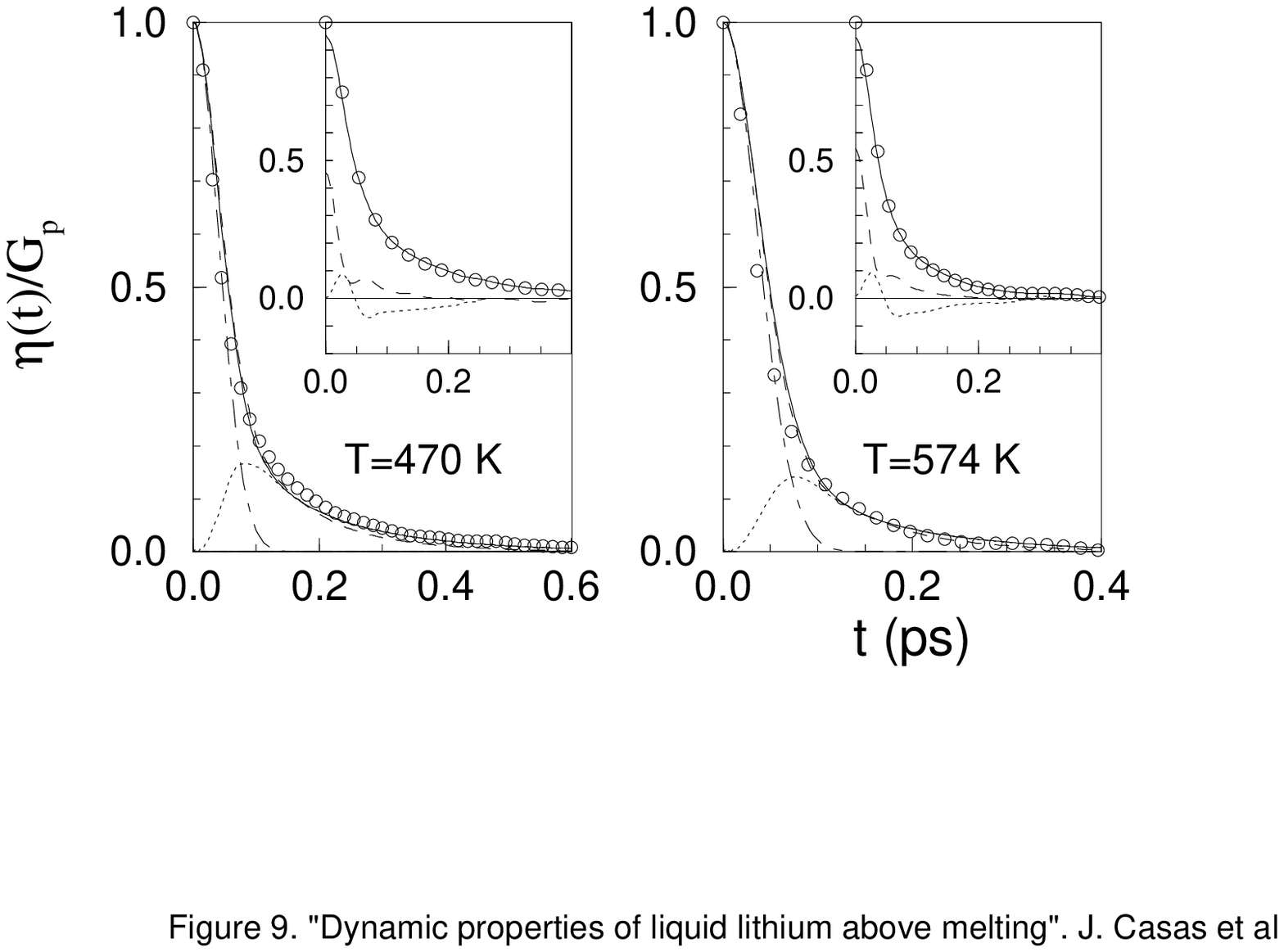,angle=-90,width=95mm}}
\end{center}
\caption{Normalised potential part of the stress 
autocorrelation function, $\eta(t)$,  
for liquid lithium at T=470 and 574 K. Open 
circles: MD results. Continuous line: theoretical results.
Dash-dotted line: binary part. Dotted line: mode-coupling component. 
Dashed line: 
viscoelastic model. The inset shows the MD results for $\eta(t)$ 
(open circles), $\eta_{pp}(t)$ (continuous line), 
10 $\times \eta_{kp}(t)$ (dotted line) 
and 10 $\times \eta_{kk}(t)$ (dashed line). } 
\label{shearTOT}
\end{figure}

Figure \ref{shearTOT} shows, for T=470 and 574 K, the MD results obtained 
for the SACF along with its three contributions. Note that, according to 
these results, the contributions from 
both the potential-kinetic, $\eta_{kp}(t)$, and 
kinetic-kinetic, $\eta_{kk}(t)$, parts of the SACF are 
negligible (less that $5\%$ of the potential-potential part) and therefore 
it justifies to restrict the theoretical calculations to the 
purely potential part of the SACF, as suggested in the theoretical framework 
presented in section \ref{shear}. 

In the figure, we have also plotted the theoretical $\eta(t)$ along with 
its binary and mode-coupling components. Whereas 
the binary part alone is quite unable to describe
the behaviour of $\eta(t)$, the inclusion of the mode-coupling
component leads to a good overall agreement with the simulation data.
However, it must be stressed that the present theoretical results  
slightly overestimate the short time ($t \leq 0.1$ ps) 
behaviour of $\eta(t)$ and 
this is mainly due to the binary component; 
more explicitly, it comes from the overestimation of the 
values for the $\tau_{\eta}$. Again, we note that the choice of the 
squared hyperbolic secant ansatz would lead to a worse description of 
the binary part.

On the other hand, the behaviour for $ t > 0.1$,  
which is completely 
determined by the mode-coupling component, is rather well described. 
The values obtained for the  
shear viscosity coefficient, $\eta$, are presented in Table 
\ref{visco}. First, we stress that the MD results compare rather well with 
the corresponding experimental values \cite{Shpil} whereas, as 
expected from figure \ref{shearTOT}, the theoretical values overestimate 
the MD ones because of their short time behaviour.

Finally, we have also plotted in figure \ref{shearTOT}, the results obtained 
by the viscoelastic approximation. It is observed that it overestimates 
the short time behaviour whereas the long-time behaviour is  
slightly underestimated; both effects counteract so as to produce 
a shear viscosity 
coefficient in good agreement with the MD results.

\section{Conclusions.}

In this paper we have evaluated several 
dynamical properties of liquid lithium at three different thermodynamic 
states. The calculations have been performed both by a theoretical framework 
and by MD simulations, using an interatomic pair potential 
derived within the Neutral 
Pseudoatom method. 

The MD results have been compared with the experimental 
dynamic structure factor, $S(q,\omega)$, obtained by IXS \cite{Sinn}, the
total dynamic 
structure factor, $S_{\rm tot}(k, \omega)$, obtained 
by INS \cite{Jong1,Jong2,Jong3}, as 
well as with the experimental data for the self-diffusion and shear viscosity 
coefficients. The overall quality of the agreement is good, supporting the 
idea that the interatomic pair potencial derived from the NPA method can be 
used to describe the dynamical properties of liquid lithium at thermodynamic 
states near the triple point. 

However, although the MD simulations can always 
be used to evaluate the dynamical properties of a system, once the interatomic 
pair potencial is given, it is also important to have a purely theoretical 
formalism which, besides of complementing the MD calculations, will allow 
the analysis of the underlying mechanisms of motion of the particles of the 
system. This is a relevant aspect of the present paper, where we   
have presented a whole theoretical framework which, by 
incorporating mode-coupling concepts, allows the self-consistent 
determination of  
single particle properties, as represented by the velocity autocorrelation 
function, its memory function, the self-diffusion coefficient and the 
self-dynamic structure factor, as well as collective properties such as 
the intermediate scattering function, the dynamic structure factor,
the autocorrelation function of the 
non-diagonal elements of the stress tensor and the shear viscosity 
coefficient. The application of this theoretical framework to the 
study of liquid lithium at thermodynamic conditions not far from the 
triple point, leads to results in overall good agreement with both 
the corresponding MD results and the experimental data. This agreement 
is particularly satisfactory at the melting point whereas some deviations 
appear at the highest temperature investigated.  

We have shown that the deficiencies in the VACF memory function, 
observed in previous works and 
related to the use of the viscoelastic model for
the intermediate scattering functions, are now significantly reduced, leading
to a much better description of the long time tail of $K(t)$.
This is basically due to the improved description of the
intermediate scattering functions, $F(k,t)$, in particular, 
of the overall amplitude of its decaying tail.
However, although the present theoretical scheme accounts for 
these features of both the single particle and the collective dynamics, 
there is still room for improvement in some dynamical magnitudes, 
like $F(k,t)$, for which we  
find some discrepancies in the amplitude of the oscillations
in the tail as compared with the corresponding MD results,
which are also reflected in the less accurate description of the side
peak of the dynamic structure factor. 

Within the present theoretical framework, a key role is played by the 
memory functions; therefore, it is important to provide an accurate 
description of these magnitudes. 
Concerning the memory functions given in equations 
(\ref{Gammatot}) and (\ref{Ktot}), this work shows that 
its short time behaviour 
is reasonably described by means of the gaussian ansatz.   
We have verified that, within the approximations made in the present work,
the hyperbolic secant squared ansatz leads to results which are too wide as
compared with the MD data.

As a final comment, we note that some authors \cite{SjoRb} have used a
simplified expresion for the relaxation time of the binary part of 
the second order memory function of the intermediate scattering functions.
Instead of using the $\tau_l(k)$ which gives the correct sixth moment of
$S(k,\omega)$, they proposed to use the simpler $\tau_s(k)$, which gives
the correct sixth moment of $S_s(k,\omega)$.
Certainly, the values of $\tau_l(k)$ oscillate around those of $\tau_s(k)$
\cite{SjoRb},
but we have found that the influence of using $\tau_s(k)$ is not 
negligible, giving a noticeably worse description of 
the intermediate scattering functions.

We end up by signaling the limitations of the formalism presented
here. First, its density/temperature range of applicability lies within the
region where the relevant slow relaxation channel is provided by the
coupling to density fluctuations, and this ceases to be valid for densities
smaller than those typical of the melting point. For these
densities, coupling to other modes, like longitudinal and/or transverse
currents, becomes increasingly important and we believe that this 
is the main reason 
for the small deviations observed in the VACF total memory function at the 
two higher temperatures studied. Moreover, we consider that an 
improvement in the description of the  
intermediate scattering function, $F(k, t)$, at all temperatures, would also 
require 
the inclusion of other modes in the corresponding second-order memory 
function. 

Second, the theory treats in a
different way the self intermediate scattering function on one hand and  
the velocity autocorrelation function and the  
intermediate scattering function on the other, since  
the mode-coupling effects are included for the latters but not for the
first. 
It would clearly be desirable a more symmetric treatment, and
further work is currently performed in that 
direction, and the results will 
be reported in due time.

\section*{Acknowledgements.}

This work has been supported by 
the DGICYT of Spain
(Project PB95-0720-CO2) and the Junta de Castilla y 
Leon (Project VA70/99). 
Dr. M.M.G. Alemany is kindly acknowledged for his help on the MD
simulations.

\newpage

\newpage

\begin{table}
\caption{Thermodynamic states studied in this work.} 
\label{thermodynamic}
\begin{tabular}{cddd}
T (K) & 470 & 526 & 574 \\
\hline
$\rho$ (\AA$^{-3}$) & 0.0445 & 0.0441 & 0.0438 \\
\end{tabular}
\end{table}

\begin{table}
\caption{Self-diffusion coefficient (in \AA$^2$/ps units),  
of liquid lithium at the thermodynamic states studied in this work. 
$D_{\rm th}$, $D_{\rm visc}$ and $D_{\rm MD}$ are the 
theoretical, viscoelastic 
and Molecular Dynamics  
results obtained in this work.}
\label{diffu}
\begin{tabular}{cddd}
T (K) & 470 & 526 & 574 \\
\hline
$D_{\rm th}$ & 0.689 & 0.875 & 1.03 \\
$D_{\rm visc}$ & 0.826 & 0.992 & 1.14 \\  
$D_{\rm MD}$ & 0.69 & 0.91 & 1.11 \\
$D_{\rm exp}$ & 0.64$\pm$0.04 \tablenotemark[1] & 
0.90$\pm$0.06 \tablenotemark[1] 
& 1.08$\pm$0.03 \tablenotemark[1] \\
 & 0.69$\pm$0.09 \tablenotemark[2] & 0.95$\pm$0.12 \tablenotemark[2] 
& 1.19$\pm$0.14 \tablenotemark[2] \\
 & 0.67$\pm$0.06 \tablenotemark[3] & 0.93$\pm$0.07 \tablenotemark[3] 
& 1.16$\pm$0.09 \tablenotemark[3] \\
\end{tabular}
\tablenotetext[1]{Ref.\  \protect\onlinecite{Seldmeier}}
\tablenotetext[2]{Ref.\  \protect\onlinecite{Lowen}}
\tablenotetext[3]{Ref.\  \protect\onlinecite{Jong1}}
\end{table}

\begin{table}
\caption{Shear viscosity (in GPa ps) of liquid lithium at the  
thermodynamic states studied in this work. $\eta_{\rm th}$ and 
$\eta_{\rm MD}$ are 
the theoretical and Molecular Dynamics results obtained in this work. }
\label{visco}
\begin{tabular}{cddd}
T (K) & 470 & 526 & 574 \\
\hline
$\eta_{\rm th}$ & 0.589 & 0.514 & 0.461 \\
$\eta_{\rm visc}$ & 0.525 & 0.481 & 0.440 \\
$\eta_{\rm MD}$ & 0.55 & 0.48 & 0.42 \\
$\eta_{\rm exp}$ & 0.57$\pm$0.03 \tablenotemark[1] & 0.50$\pm$0.03 
\tablenotemark[1] 
& 0.45$\pm$0.03 \tablenotemark[1] \\
\end{tabular}
\tablenotetext[1]{Ref.\  \onlinecite{Shpil}} 
\end{table}

\end{document}